# Low-Complexity Acoustic Scene Classification Using Data Augmentation and Lightweight ResNet

Yanxiong Li, Wenchang Cao, Wei Xie, Qisheng Huang, Wenfeng Pang, Qianhua He

*School of Electronic and Information Engineering, South China University of Technology, Guangzhou, China*

eeyxli@scut.edu.cn, wenchangcao98@163.com, chester.w.xie@gmail.com, 839508665@qq.com, wenfengpang@gmail.com, eeqhhe@scut.edu.cn

*Abstract*—We present a work on low-complexity acoustic scene classification (ASC) with multiple devices, namely the subtask A of Task 1 of the DCASE2021 challenge. This subtask focuses on classifying audio samples of multiple devices with a low-complexity model, where two main difficulties need to be overcome. First, the audio samples are recorded by different devices, and there is mismatch of recording devices in audio samples. We reduce the negative impact of the mismatch of recording devices by using some effective strategies, including data augmentation (e.g., mix-up, spectrum correction, pitch shift), usages of multi-patch network structure and channel attention. Second, the model size should be smaller than a threshold (e.g., 128 KB required by the DCASE2021 challenge). To meet this condition, we adopt a ResNet with both depthwise separable convolution and channel attention as the backbone network, and perform model compression. In summary, we propose a low-complexity ASC method using data augmentation and a lightweight ResNet. Evaluated on the official development and evaluation datasets, our method obtains classification accuracy scores of 71.6% and 66.7%, respectively; and obtains Log-loss scores of 1.038 and 1.136, respectively. Our final model size is 110.3 KB which is smaller than the maximum of 128 KB.

## I. Introduction

Acoustic scene classification (ASC) aims to classify each audio sample into one category of pre-defined acoustic scenes. As one critical task in the challenge on Detection and Classification of Acoustic Scenes and Events (DCASE) [1], ASC has recently become one of the hottest topics in the field of audio and acoustic signal processing [2]-[26]. With the development of the study on the ASC problem, some state-of-the-art techniques are proposed to solve the ASC problem.

In this work, we concentrate on solving the problem of low-complexity ASC with multiple devices which is the subtask A of Task 1 of the DCASE2021 challenge [27]. This task requires to classify audio samples recorded by many devices (both real and simulated) through a low-complexity model. To deal with the mismatch problem of recording devices in audio samples, many researchers have proposed many methods which are mainly divided into two types: data processing and model building.

The first type focuses on front-end data processing. Audio samples acquired by various devices generally have different time-frequency properties. One common practice for alleviating the negative impact of recording device mismatch is to collect as many audio samples as possible from different recording devices, but it requires a lot of manpower. Besides, more audio samples can be obtained by some low-cost and effective ways, such as spectrum correction, frequency shift, mix-up [28], [29].

The second type focuses on back-end model building, where the model is designed to have the abilities to distinguish different types of acoustic scenes and to avoid the influence of different devices. The methods based on the models of convolutional neural network (CNN) and the CNN's variants are dominant solutions for ASC tasks [4]-[9], [30]-[36]. The CNN has powerful ability to learn discriminative information from its input spectrogram which is one of the most effective and widely used audio features in ASC task. In addition, some researchers modify existing networks and apply them to the ASC task with satisfactory results. For example, McDonnell et al separate the ResNet into two pathways, and the separation of high and low frequencies is proved to be effective for enhancing the network's adaptability to multi-device audio samples [33]. Koutini et al explore the influence of receptive field on the performance of the ASC methods [34]. Su et al modify the Xception network to build their model for making prediction [35].

The works above have promoted the development of ASC, but there are still shortcomings when they are used to tackle the problem of low-complexity ASC. For example, the model size in these methods is very large, and these methods are not robust to multiple devices. Inspired by the successes of data augmentation and CNN-based models for audio and vision classification in previous works, we propose a method for low-complexity ASC using data augmentation techniques and a lightweight ResNet with channel attention (CA). In our method, depthwise separable convolution (DSC) [37] rather than standard convolution is used as the basic module for building a lightweight ResNet, and the CA is adopted to utilize channel information after DSC. In addition, some techniques of data augmentation are utilized to mitigate the negative influence of recording devices mismatch on the proposed method. Experimental results have proved the effectiveness of our method. In short, the contributions of the work in this paper are summarized as follows.

1) We propose a low-complexity ASC method using data augmentation techniques and a lightweight network with CA.

2) We discuss contributions of main modules of our method (e.g., CA, data augmentation), and compare the proposed method with baseline method on public and official datasets.

## II. Method

As depicted in Fig. 1, the framework of the proposed method consists of two parts: data augmentation and model building.

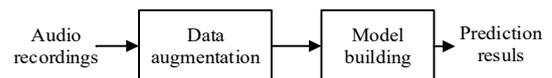

Fig. 1: *The framework of the propoised method.*

### A. Data Augmentation

To reduce the negative impact of the recording device mismatch on the performance of our method, some data augmentation techniques are adopted in our method. These techniques are described as follows.

The first technique is mix-up which is theoretically simple but effective [29]. It makes our model behave consistently to training samples from different devices. This modeling technique can enhance robustness of the model on training samples. The mix-up algorithm adopted in our paper is theoretically consistent with the original one. In practical implementation, our practice for mix-up is slightly different from original one in [29]. In each time, we load two batches of audio samples instead of two audio samples (used by [29]), into memory and then process them by $X' = \lambda X_i + (1-\lambda)X_j$. $X_i$ and $X_j$ are two batches of audio samples from the shuffled training data, and $\lambda$ follows Beta distribution, i.e. $\lambda \sim \text{Beta}(\alpha, \alpha)$ and $\alpha$ is experimentally set to 0.4. Through loading two batches of audio samples, more information can be used during the procedure of mixing, which is of benefit to the performance improvement of our method.

The second technique is spectrum correction which shows moderate device adaptation properties [38]. Here, the spectrum correction is adjusted before it is applied to ASC. It aims to convert the input spectrum to a corrected spectrum by one ideal device (i.e., reference device). Its implementation is composed of two stages. First, a correction coefficient is obtained by computing the mean of $n$ pairs of aligned spectra. Here, the value of $n$ is experimentally set to 150. The correction coefficient of device $A$ to the reference device is defined as the ratio of the frequency response of the reference device to the counterpart of device $A$. The frequency response of the reference device is the mean of the counterparts of several devices. In the second stage, the corrected spectrum of device $A$ is produced through multiplying the correction coefficient with the original spectrum of the audio samples recorded by device $A$.

The third technique is pitch shift [39], which is to resample the original audio samples at various sampling frequencies with one specific step size. The pitches of audio samples acquired by various devices are usually different to each other, which is beneficial for ASC. Therefore, pitch shifting is carried out before extracting the log-Mel spectrogram (LMS) from each audio sample.

The fourth technique is audio-mix which is inspired by the work in [40]. We mix two audio samples that are randomly selected from the same kind of acoustic scenes for simulating more recording devices, smoothing transition among various recording devices, and reducing differences among different recording devices.

*B. Model Building*

Our model is a lightweight ResNet which inherits merits from MobileNet [41] and ResNet [42], whose main module is inverted residual blocks with DSC and CA. The motivation for the adoption of both DSC and CA is based on two reasons. Firstly, the DSC can remarkably decrease the model's size. Secondly, the CA can make the model concentrate on the critical channel information which is not efficiently used during the operation of DSC. Hence, the integration of the DSC and CA into the inverted residual blocks is expected to be able to make our model more concise with better generalization ability for ASC with multiple devices, which is also one technical contribution of this work.

*Framework of Our Model*

As shown in Fig. 2, our model is a deep architecture consisting of two pathways of inverted residual blocks with DSC and CA, maximum and global average pooling layers, fully-connected and Softmax layers. The motivation for the usage of two pathways of inverted residual blocks is that deep feature maps learned by two pathways of inverted residual blocks are expected to be more effective for representing the differences of time-frequency properties among various acoustic scenes and thus produce a better result compared to one pathway of inverted residual blocks.

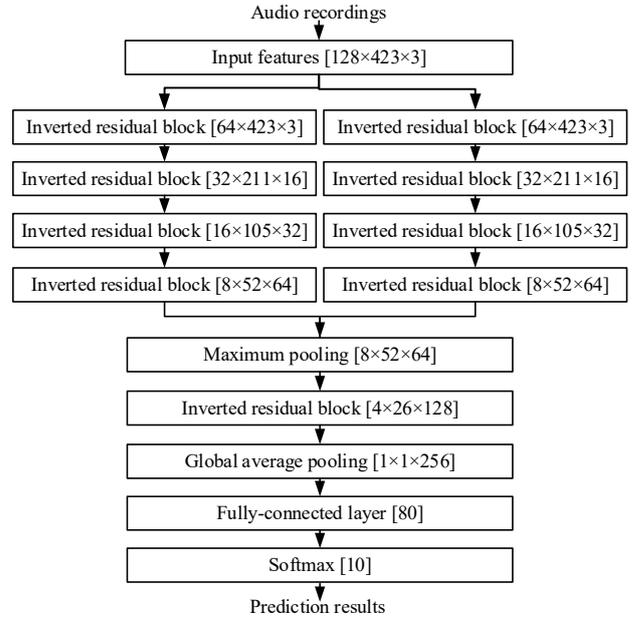

Fig. 2: *Framework of our model.*

Our model's structure and parameters are shown in Fig. 2. The three digits in the square bracket, such as 32, 211 and 16 in [32×211×16], represent height of the feature map, width of the feature map, and number of channels, respectively. The workflow of our model is as follows. First, input features are extracted from audio samples and then are split into two subsets along the first dimension (the dimension of filter-bank) of input features. Next, the two subsets of input features are fed to two pathways of inverted residual blocks in parallel. Afterwards, the outputs of these two pathways of inverted residual blocks are concatenated along the first dimension of input features and then their concatenation is sequentially fed to one maximum pooling layer, one inverted residual block and one global average pooling layer for further transformation. Finally, the transformed feature map is flattened by a fully-connected layer and then fed to a Softmax layer to make a prediction of acoustic scene for each input audio sample.

*Inverted Residual Block*

The framework of one inverted residual block is depicted in Fig. 3, which mainly includes the layers of DSC, batch normalization, ReLU, channel attention, and Maximum pooling.

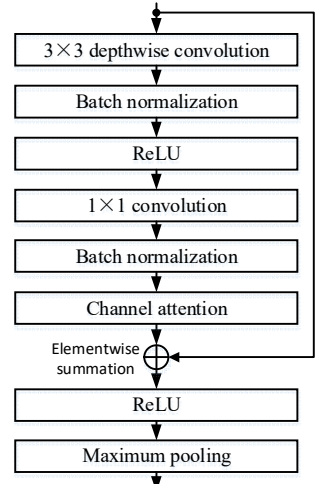

Fig. 3: *Framework of one inverted residual block.*

The inverted residual block with DSC and CA is the novel part of our model. One convolutional kernel of the DSC is responsible for one channel, and one channel is convoluted by only one convolutional kernel. The DSC is implemented via decomposing the standard convolution into two steps: depthwise convolution and 1×1 pointwise convolution [37]. The depthwise convolution applies one filter to each input channel, while the pointwise convolution applies 1×1 convolution to combine the outputs of various depthwise convolutions.

*Channel Attention*

A channel attention mechanism [43] is introduced to our model, whose framework is depicted in Fig. 4. The dimension of input feature $F$ is $H×W×C$, where $H$, $W$ and $C$ represent height of the feature map, width of the feature map, and number of channels, respectively. First, the operations of maximum pooling (MaxPool) and average pooling (AvgPool) are executed on the input feature $F$ for generating two channel descriptions with dimension of $1×1×C$. Then, these two channel descriptions are independently fed to a multi-layer perceptron (MLP) for obtaining their respective transformed features. That is, this MLP is shared for these two channel descriptions. Afterwards, the weight coefficient $M_C$ is obtained by element-wisely summing these two transformed features from the MLP followed by a Sigmoid activation function. Finally, a new scaled feature $F'$ is produced via multiplying the weight coefficient $M_C$ with the feature $F$.

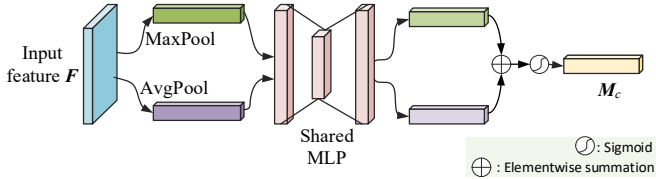

Fig. 4: *Framework of channel attention*.

### III. EXPERIMENTS

In this section, we will present experimental data, setup, results, and discussions in detail.

*A. Experimental Data*

Experimental data consists of development dataset and evaluation dataset and has 10 types of acoustic scenes, which is available for research purpose on the website of the DCASE2021 challenge [44]. The development dataset consists of the training and validation subsets. The training subset and the validation subset have 13962 and 2968 audio samples, respectively. The development dataset contains audio samples from ten cities and nine devices: three real devices (*A*, *B*, *C*) and six simulated devices (*S*1-*S*6). Audio samples of devices *B*, *C*, and *S*1-*S*6 are randomly chosen from the audio segments of simultaneous recordings. Audio samples of devices *B*, *C*, and *S*1-*S*6 overlap with the audio samples of device *A*, without overlapping with each other. The length of audio samples in the development dataset is 64 hours in total. In the development dataset, the data proportions of training and validation subsets are 70% and 30%, respectively. The evaluation dataset contains audio samples from twelve cities, eleven devices. There are six new devices (not in the development dataset): a real device *D* and simulated devices *S*7-*S*11. The evaluation dataset has 22 hours of audio samples.

*B. Experimental Setup*

All audio samples are saved as: 44.1 kHz sampling rate, 16 bits quantization, and mono channel. First, audio samples are split into audio frames by a hamming window whose duration is 2048 with 50% overlapping. Afterwards, short-time Fourier transform is performed on each audio frame for producing linear power spectrum which is smoothed with a bank of triangular filters for obtain LMS. The center frequencies of all triangular filters are uniformly distributed on the Mel-scale. Besides, to enhance the discriminative capability of audio feature, the delta, and delta-delta coefficients of the LMS are derived too. The LMS, its delta and delta-delta coefficients are concatenated along the channel axis and adopted as input feature whose size is: 128×423×3, where 128, 423 and 3 denote the numbers of filter-bank, frame and channel, respectively. The three channels are the LMS, the delta coefficients and the delta-delta coefficients of the LMS.

The baseline method is a combination of a CNN and a feature of log Mel-band energies. The CNN used in the baseline method is composed of three convolutional layers and a fully-connected layer. The details of the baseline method are referred to [44]. The major difference between our method and the baseline method is the back-end model. That is, our model mainly consists of two pathways of inverted residual blocks with DSC and CA, whereas the model in the baseline method is a common CNN with standard convolution without DSC and CA.

All experiments are done using the toolkit of Keras. Adam is used as the optimizer, and categorical cross entropy is adopted as the loss function. Classification accuracy (the higher the better) and log-loss (the lower the better) are used as performance metrics. Adam's initial value is set to 0.001, and cosine annealing algorithm is adopted as the learning strategy. There is a cosine relationship between training epochs and learning rate. The maximum and minimum learning rates are set to $10^{-3}$ and $10^{-7}$, respectively. The batch size is set to 16. The checkpoint with the largest validation accuracy is used as the best model. The models are trained using 3 different seeds and choose the average accuracy as final result.

In subtask A of Task 1 of DCASE2021 challenge, the maximum of the model size is 128 KB excluding parameters with zero-values. In other words, the model contains at most 32768 parameters when each parameter is represented by 32-bit floating numbers (i.e., 32768 parameters × 32 bits/parameter = 128 × 8 bits/Byte × 1024 Bytes = 128 KB). To meet this requirement, besides the adoption of lightweight ResNet with DSC and CA, a quantization operation is performed for further compressing our trained model. The quantization method is from the Tensorflow2 [43] which is adopted to convert the 32-bit floating parameters of our model into the 16-bit floating parameters. Specifically, floating-point numbers are saved by scientific counting method. During the process of converting 32-bit to 16-bit floating numbers, the sign and exponential bits remain unchanged, and the first half part of the numeric bits of each 32-bit floating number are used as numeric bits of the 16-bit floating number.

*C. Experimental Results*

Table I lists the contribution of CA to our and baseline methods regarding classification accuracy on the development dataset without data augmentation. Classification accuracies of two methods are upgraded after adopting CA. For example, our method with CA obtains classification accuracy improvement by 1.8% (67.2% - 65.4%), compared to our method without CA.

TABLE I. CLASSIFICATION ACCURACIES OBTAINED BY OUR METHOD AND BASELINE METHOD WITH OR WITHOUT CA

|  | Baseline | Ours |
|---|---|---|
| without channel attention | 49.8% | 65.4% |
| with channel attention | 51.3% | 67.2% |

Table II shows the contributions of various data augmentation techniques to the increase of classification accuracies for our and baseline methods on the development dataset. Classification accuracies of these two methods are increased after introducing data augmentation techniques, and every data augmentation technique contributes to performance improvement of these two methods. Moreover, our and baseline methods obtain the largest classification accuracies of 71.4% and 54.9%, respectively, when all data augmentation techniques are adopted. For example, our method with mix-up and spectrum correction techniques obtains classification accuracy improvement by 1.9% (69.1% - 67.2%) and 2.4% (69.6% - 67.2%), respectively, compared with the methods without any data augmentation techniques. Among these data augmentation techniques, pitch shift technique produces the highest improvement by 3.3% (70.5% - 67.2%) for our method, while the technique of mix audio generates the highest improvement by 3.2% (54.5% - 51.3%) for the baseline method. The reasons why pitch shift and mix audio are more effective than other techniques are that larger noise interferences are integrated into the original audio samples, and these noise interferences enlarge the differences of time-frequency properties among the audio samples of different recording devices. Hence, they can produce higher performance gain.

TABLE II. CLASSIFICATION ACCURACIES OBTAINED BY OUR METHOD AND BASELINE METHOD WITH OR WITHOUT DATA AUGMENTATION TECHNIQUES

| Data augmentation techniques | Baseline | Ours |
|---|---|---|
| None (without any techniques) | 51.3% | 67.2% |
| Mix-up | 52.4% | 69.1% |
| Spectrum correction | 53.1% | 69.6% |
| Pitch shift | 54.2% | 70.5% |
| Mix audio | 54.5% | 69.9% |
| All above | 54.9% | 71.4% |

Table III shows the results of our and baseline methods on the development dataset. The number of non-zero parameters and the model size of our method are larger than the counterparts of baseline method, and the margins of the number of non-zero parameters and model size between the baseline method and our method are 8800 (56486 - 47686) and 20 KB (110.3 -90.3 KB), respectively. Our method is heavier than the baseline method regarding both the number of non-zero parameters and model size. Our method exceeds the baseline method in terms of classification accuracy. The margin of classification accuracy between the baseline and our methods is 16.5%. In addition, our model size after quantization compression is 110.3 KB (56486 parameters × 16 bits/parameter = 110.3 × 8 bits/Byte × 1024 Bytes = 110.3 KB) which is smaller than the limit of 128 KB.

TABLE III. RESULTS OF BASELINE AND OUR METHODS

| Model | Accuracy | Number of non-zero parameters | Model size(KB) |
|---|---|---|---|
| Baseline | 54.9% | 47686 | 90.3 |
| Ours | 71.4% | 56486 | 110.3 |

To know confusions among various scenes, Fig. 5 shows the confusion matrix obtained by our method on development dataset.

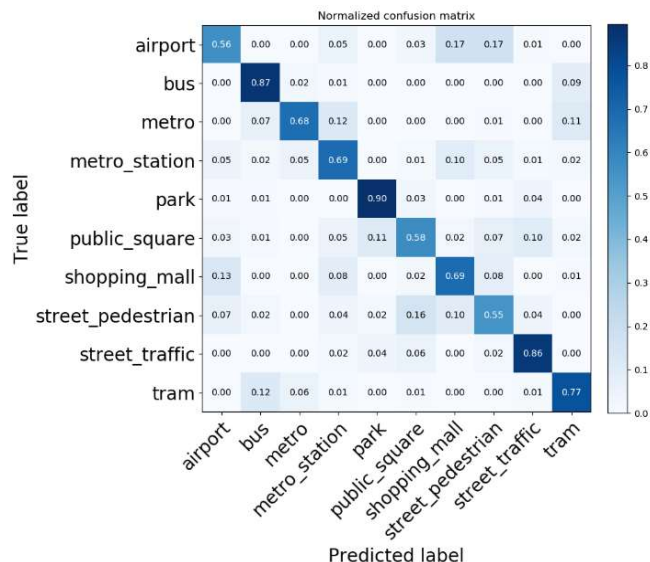

Fig. 5: *Confusion matrix of our method on the development dataset.*

The accuracies for *Public square*, *Airport*, and *Street pedestrian* are lower than that for other acoustic scenes. The reason is that these three acoustic scenes have larger intra-class differences. In addition, the confusions between *Airport* and *Shopping mall* are larger than that between other pairs of acoustic scenes. The reason is that the time-frequency characteristics of these two acoustic scenes are more similar. As a result, they are more easily confused with each other.

Finally, we evaluate our method on the evaluation dataset. It should be noted that the labels of the evaluation dataset are not available and the results in Table IV are given by the organizer of the DCASE2021 challenge [44]. As presented in Table IV, our method obtains Log-loss score of 1.136 and classification accuracy score of 66.7%, and exceeds the baseline method in terms of Log-loss and classification accuracy.

TABLE IV. RESULTS OBTAINED BY OUR METHOD AND BASELINE METHOD ON THE EVALUATION DATASET

| Acoustic scenes | Baseline | | Ours | |
|---|---|---|---|---|
| | CA (%) | Log-loss | CA (%) | Log-loss |
| Airport | 24.0 | 2.077 | 43.7 | 1.461 |
| Bus | 44.6 | 1.615 | 77.0 | 1.007 |
| Metro | 54.4 | 1.159 | 67.2 | 1.169 |
| Metro station | 37.8 | 1.955 | 59.3 | 1.343 |
| Park | 52.7 | 2.173 | 84.1 | 0.753 |
| Public square | 24.4 | 2.455 | 49.9 | 1.485 |
| Shopping mall | 63.8 | 1.227 | 77.3 | 1.006 |
| Street pedestrian | 39.9 | 1.744 | 41.9 | 1.576 |
| Street traffic | 56.4 | 1.825 | 85.0 | 0.618 |
| Tram | 58.1 | 1.073 | 81.6 | 0.937 |
| Average | 45.6 | 1.730 | 66.7 | 1.136 |

## IV. CONCLUSIONS

We presented our work on the low-complexity ASC with multiple devices. We proposed a method by a lightweight ResNet with DSC and CA. The whole network's structure and some strategies designed in this paper are novel. For example, the DSC is adopted to make our network light enough, and the CA is integrated into the inverted residual block to exchange channel information. In addition, considering the differences of time-frequency properties among different frequency bands of audio feature, we first divide the input audio feature into two parts: low-frequency and high-frequency parts along the dimension of frequency axis, and then fuse them.

Our method exceeded baseline method in terms of Log-loss and classification accuracy. Moreover, the size of our model is 110.3KB, and is lower than the limit of 128 KB. Our method is effective for tackling the problem of low-complexity ASC. Limitations of our method are that front-end data augmentation and back-end model building are separately instead of jointly implemented, and that only one model instead of multi-model combination is adopted. The next works include exploring other model structures, designing the techniques of data augmentation, jointly optimizing both front-end data augmentation and back-end model building, and assembling models.

ACKNOWLEDGMENT

This work was supported by national natural science foundation of China (62111530145, 61771200), international scientific research collaboration project of Guangdong Province (2021A0505030003), and Guangdong basic and applied basic research foundation (2021A1515011454, 2022A1515011687).

REFERENCES

[1] T. Heittola, A. Mesaros, and T. Virtanen, "Acoustic scene classification in DCASE 2020 challenge: generalization across devices and low complexity solutions," in *Proc. of DCASE Workshop*, 2020.

[2] Y. Li, X. Li, Y. Zhang, W. Wang, M. Liu, and X. Feng, "Acoustic scene classification using deep audio feature and BLSTM network," in *Proc. of ICALIP*, 2018, pp. 371-374.

[3] S. Singh, H. L. Bear, and E. Benetos, "Prototypical networks for domain adaptation in acoustic scene classification," in *Proc. of IEEE ICASSP*, 2021, pp. 346-350.

[4] Y. Liu, A. Neophytou, S. Sengupta, and E. Sommerlade, "Cross-modal spectrum transformation network for acoustic scene classification," in *Proc. of IEEE ICASSP*, 2021, pp. 830-834.

[5] H. Hu et al., "A two-stage approach to device-robust acoustic scene classification," in *Proc. of IEEE ICASSP*, 2021, pp. 845-849.

[6] Y. Li, M. Liu, W. Wang, Y. Zhang and Q. He, "Acoustic scene clustering using joint optimization of deep embedding learning and clustering iteration," *IEEE TMM*, vol. 22, no. 6, pp. 1385-1394, 2020.

[7] Z. Ren, Q. Kong, J. Han, M. D. Plumbley, and B. W. Schuller, "CAA-Net: conditional atrous CNNs with attention for explainable device-robust acoustic scene classification," *IEEE TMM*, 2020.

[8] S. S. R. Phaye, E. Benetos, and Y. Wang, "SubSpectralNet - using sub-spectrogram based convolutional neural networks for acoustic scene classification," in *Proc. of IEEE ICASSP*, 2019, pp. 825-829.

[9] S. Takeyama, T. Komatsu, K. Miyazaki, M. Togami, and S. Ono, "Robust acoustic scene classification to multiple devices using maximum classifier discrepancy and knowledge distillation," in *Proc. of EUSIPCO*, 2021, pp. 36-40.

[10] A. I. Mezza, E. A. P. Habets, M. Müller, and A. Sarti, "Unsupervised domain adaptation for acoustic scene classification using band-wise statistics matching," in *Proc. of EUSIPCO*, 2021, pp. 11-15.

[11] V. Abrol, and P. Sharma, "Learning hierarchy aware embedding from raw audio for acoustic scene classification," *IEEE/ACM TASLP*, vol. 28, pp. 1964-1973, 2020.

[12] M. C. Green, S. Adavanne, D. Murphy, and T. Virtanen, "Acoustic scene classification using higher-order ambisonic features," in *Proc. of IEEE WASPAA*, 2019, pp. 328-332.

[13] L. Pham, I. McLoughlin, H. Phan, R. Palaniappan, and A. Mertins, "Deep feature embedding and hierarchical classification for audio scene classification," in *Proc. of IJCNN*, 2020, pp. 1-7.

[14] Y. Wu, and T. Lee, "Time-frequency feature decomposition based on sound duration for acoustic scene classification," in *Proc. of IEEE ICASSP*, 2020, pp. 716-720.

[15] K. Drossos, P. Magron, and T. Virtanen, "Unsupervised adversarial domain adaptation based on the Wasserstein distance for acoustic scene classification," in *Proc. of IEEE WASPAA*, 2019, pp. 259-263.

[16] C. Paseddula, and S. V. Gangashetty, "Acoustic scene classification using single frequency filtering cepstral coefficients and DNN," in *Proc. of IEEE IJCNN*, 2020, pp. 1-6.

[17] X. Bai, J. Du, J. Pan, H.S. Zhou, Y.H. Tu, and C.H. Lee, "High-resolution attention network with acoustic segment model for acoustic scene classification," in *Proc. of IEEE ICASSP*, 2020, pp. 656-660.

[18] Y. Li, W. Cao, K. Drossos, and T. Virtanen, "Domestic activity clustering from audio via depthwise separable convolutional autoencoder network," in *Proc. of IEEE MMSP*, 2022, pp. 1-6. Online: https://arxiv.org/ftp/arxiv/papers/2208/2208.02406.pdf

[19] W. Xie, Q. He, Z. Yu, and Y. Li, "Deep mutual attention network for acoustic scene classification," *Digital Signal Processing*, vol.123, 2022.

[20] L. Pham, H. Phan, T. Nguyen, R. Palaniappan, A. Mertins, and I. McLoughlin, "Robust acoustic scene classification using a multi-spectrogram encoder-decoder framework," *Digital Signal Processing*, vol. 110, 2021.

[21] J.-w. Jung, H.-j. Shim, J.-h. Kim, and H.-J. Yu, "DCASENET: An integrated pretrained deep neural network for detecting and classifying acoustic scenes and events," in *Proc. of IEEE ICASSP*, 2021, pp. 621-625.

[22] C. Paseddula, and S.V. Gangashetty, "Late fusion framework for acoustic scene classification using LPCC, SCMC, and log-Mel band energies with deep neural networks," *Applied Acoustics*, vol. 172, 2021.

[23] M.A. Alamir, "A novel acoustic scene classification model using the late fusion of convolutional neural networks and different ensemble classifiers," *Applied Acoustics*, vol. 175, 2021.

[24] K. Imoto, "Acoustic scene classification using multichannel observation with partially missing channels," in *Proc. of EUSIPCO*, 2021, pp. 875-879.

[25] S. Song, B. Desplanques, C. De Moor, K. Demuynck, and N. Madhu, "Robust acoustic scene classification in the presence of active foreground speech," in *Proc. of EUSIPCO*, 2021, pp. 995-999.

[26] J. Zhao, Q. Kong, X. Song, Z. Feng, and X. Wu, "Feature alignment for robust acoustic scene classification across devices," in *IEEE Signal Processing Letters*, 2022, doi: 10.1109/LSP.2022.3145336.

[27] I. Martín-Moraló, T. Heittola, A. Mesaros, and T. Virtanen, "Low-complexity acoustic scene classification for multi-device audio: analysis of DCASE 2021 challenge systems," 2021.arXiv:2105.13734.

[28] M. Kośmider, "Calibrating neural networks for secondary recording devices," in *Tech. Rep of DCASE2019 Challenge*.

[29] H. Zhang, M. Cisse, Y. N. Dauphin, and D. LopezPaz, "MixUp: Beyond empirical risk minimization," in *Proc. of ICLR*, 2018, pp.1-13.

[30] D. Barchiesi, D. Giannoulis, D. Stowell, and M. D. Plumbley, "Acoustic scene classification: Classifying environments from the sounds they produce," *IEEE SPM*, vol. 32, no. 3, pp. 16-34, 2015.

[31] H.K. Chon, Y. Li, W. Cao, Q. Huang, W. Xie, W. Pang, and J. Wang, "Acoustic scene classification using aggregation of two-scale deep embeddings," in *Proc. of IEEE ICCT*, 2021, vol. 4, pp. 1341-1345.

[32] J. Salamon and J.P. Bello, "Deep convolutional neural networks and data augmentation for environmental sound classification," *IEEE SPL*, vol. 24, no. 3, pp. 279-283, 2017.

[33] M.D. McDonnell and W. Gao, "Acoustic scene classification using deep residual networks with late fusion of separated high and low frequency paths," in *Proc. of IEEE ICASSP*, 2020, pp. 141-145.

[34] K. Koutini, H. Eghbal-zadeh, G. Widmer, and J. Kepler, "CP-JKU submissions to DCASE'19: Acoustic scene classification and audio tagging with receptive-field-regularized CNNs," in *DCASE Challenge*, 2019, pp. 1-5.

[35] Y. Su, K. Zhang, J. Wang, and K. Madani, "Environment sound classification using a two-stream CNN based on decision-level fusion," *Sensors*, vol. 19, no. 7, pp. 1-15, 2019.

[36] Y. Zeng, Y. Li, Z. Zhou, R. Wang, and D. Lu, "Domestic activities classification from audio recordings using multi-scale dilated depthwise separable convolutional network," in *Proc. of IEEE MMSP*, Tampere, Finland, 2021, pp. 1-5.

[37] K. Drossos, S.I. Mimilakis, S. Gharib, Y. Li, and T. Virtanen, "Sound event detection with depthwise separable and dilated convolutions," in *Proc. of IJCNN*, 2020, pp. 1-7.

[38] T. Nguyen, F. Pernkopf, and M. Kosmider, "Acoustic scene classification for mismatched recording devices using heatedup Softmax and spectrum correction," in *Proc. of IEEE ICASSP*, 2020, pp. 126-130.

[39] S. Watanabe, and M. Tsuzaki, "Pitch shifts by the overlap of identical pulse trains with a delay and its relation to the binaural differences," *Acoustical Science and Technology*, vol. 41, no. 1, pp. 433-434, 2020.

[40] H. Hu, C.H.H. Yang, X. Xia, et al, "Device-robust acoustic scene classification based on two-stage categorization and data augmentation," in *Tech. Rep of DCASE2020 Challenge*.

[41] M. Sandler, A. Howard, M. Zhu, A. Zhmoginov, and L.C. Chen, "MobileNetV2: inverted residuals and linear bottlenecks," in *Proc. of IEEE CVPR*, 2018, pp. 4510-4520.

[42] K. He, X. Zhang, S. Ren, and J. Sun, "Deep residual learning for image recognition," in *Proc. of IEEE CVPR*, pp. 770-778, 2016.

[43] S. Woo, J. Park, J.Y. Lee, I.S. Kweon, "CBAM: convolutional block attention module," in *Proc. of ECCV*, 2018, pp. 3-19.

[44] http://dcase.community/challenge2021/task-acoustic-scene-classification.